\documentclass[epsfig]{mn2e}
\usepackage{psfig}
\usepackage{mnras_cite}
\newcommand{\msun}{{\rm M_{\odot}}}

\newlength{\tskip}
\newlength{\colwidth}

\newcommand{\beq}{\begin{equation}}
\newcommand{\eeq}{\end{equation}}
\newcommand{\beqa}{\begin{eqnarray}}
\newcommand{\eeqa}{\end{eqnarray}}

\newcommand{\ed}{{\rm d}}

\begin{document}

\title[First Sources in Infrared Light]{First Sources in 
Infrared Light: Stars, Supernovae and Miniquasars}
\author[Cooray \& Yoshida]{Asantha Cooray$^1$ and Naoki Yoshida$^{2,3}$\\
$^1$Theoretical Astrophysics, California Institute of Technology, 
MS 130-33, Pasadena, CA 91125, USA\\
$^2$Division of Theoretical Astrophysics, National Astronomical 
Observatory Japan, Mitaka, Tokyo 181-8588, Japan\\
$^3$Department of Physics and Astrophysics, Nagoya University, 
Nagoya 464-8602, Japan}

\maketitle

\begin{abstract}
The cosmic infrared background (IRB) at wavelengths between 1 $\mu$m 
and 3 $\mu$m provides a useful probe of early star-formation prior to 
and during reionization. To explain the high optical depth to electron 
scattering, as measured by the Wilkinson Microwave Anisotropy Probe (WMAP), 
one requires significant star-formation activity at redshifts 10 and higher.
In addition to massive stars, the IRB flux may be contributed by a population of early miniquasars. 
We study the relative contributions from first stars, supernovae and 
quasars to the IRB for reasonable star formation rates at high redshift. 
If miniquasars radiate efficiently at the Eddington-limit, 
current background measurements limit the fraction of mass in first stars 
that is converted to seed black holes to be roughly less than 10\%.
In the case of supernovae, though an individual supernova is much brighter 
than the progenitor star, due to the shorter lifetime of order few months, 
the fractional contribution to the IRB remains at a level of 
10\% and below when compared to the same contribution from stars. 
The bright supernovae may, however, be directly detectable
by future large ground-based and space telescopes.  
\end{abstract}


\section{Introduction}

The Wilkinson Microwave Anisotropy Probe (WMAP) has provided strong 
evidence for an optical depth for electron scattering of 0.17 $\pm$ 0.04 
based on the large scale polarization pattern related to rescattering of 
Cosmic Microwave Background (CMB) photons (Kogut et al. 2003). 
If the reionization process is described as instantaneous and homogeneous,
the measured optical depth implies a reionization redshift of 
$\sim 17 \pm 5$ in a spatially flat universe.
Interestingly, the derived redshift for reionization is at the high end 
of expectations in models in which stellar populations are dominant
reionization sources (e.g., Cen 2003; Fukugita \& Kawasaki 2003;
Venkatesan, Tumlinson \& Shull 2003; Wyithe \& Loeb 2003; 
Sokasian et al. 2004). Thus, in some of these models, primordial stars, 
the so-called Population III stars, are assumed to be rather massive 
and hence have a large UV photon emission rate,
as suggested by recent theoretical studies 
(e.g. Abel, Bryan \& Norman 2002; Bromm, Coppi \& Larson 2002; 
Bromm, Kudritzki \& Loeb 2001; Schaerer 2002, 2003).

The formation of such massive stars in the early universe has
many important cosmological implications (e.g., Carr, Bond, \& Arnett 1984; 
Oh, Cooray \ Kamionkowski 2003; Yoshida, Bromm \& Hernquist 2004).
Because radiation below the Lyman-limit is absorbed by neutral hydrogen, 
luminous sources at redshifts 10 and higher are generally seen only in the 
near-IR band. In particular, if Pop~III stars are formed primarily at 
redshifts between 10 and 30, they are expected to contribute to the infrared 
background (IRB) light at wavelengths between 1 and 5 $\mu$m at present
epoch (Bond, Carr \& Hogan 1986). Recent estimates based on theoretical models 
suggest that a large fraction of the IRB total intensity may indeed be due to 
these stars (Santos et al. 2002; 
Salvaterra \& Ferrara 2003; Cooray et al. 2004). 
A substantial IRB could arise from the Pop~III stars not only due 
to the direct emission associated with these stars, but also due to 
indirect processes that lead to free-free and Lyman-$\alpha$ emission from 
the ionized nebulae, or H{\sc ii} regions, surrounding these stars.

Observationally, about 20-40\% of the near-IR flux (between $\sim$ 1 $\mu$m 
and 3 $\mu$m) has been resolved by point sources (Cambr\'esy et al. 2001; 
Totani et al. 2001); while J- and K-band source counts go down to
AB-magnitudes of $\sim$ 28 and 25, respectively, the cumulative surface 
brightness converges by a magnitude around 23, say, in the K-band (see for 
a review of existing data by Pozzetti \& Madau 2000).  
The missing IR flux could be either due to a high-z population of galaxies, 
such as the proto-galaxy population related to first stars (Santos et al. 2002; 
Salvaterra \& Ferrara 2003), or a population of low surface brightness galaxies 
at nearby redshifts. While there still remains uncertainties, one can potentially 
understand the presence of first luminous sources and some details related to 
their population, such as the number density, by characterizing the background 
light from optical to infrared at wavelengths of a few micron.
In this wavelength range, final products of these massive stars are also 
expected to contribute to IRB, in addition to the direct emission from stars; 
Radiation from supernovae and primordial black holes, in the form of miniquasars, 
may be significant sources of emission. The role of final products in contributing 
to the IRB, relative to the progenitor stars, depends on the initial mass function 
of the first stars. 

As the end state of a massive star above $\sim 260 \msun$ is a complete collapse 
to a black hole (Heger \& Woosley 2002), if the first stellar population is extremely 
top-heavy with all mass above this limit, then one expects roughly a similar number 
density of black holes as that of stars. These black holes grow via accretion and/or 
mergers and may partly be responsible for early reionization if they radiate as 
miniquasars (Haiman \& Loeb 1998; Madau et al. 2004; Ricotti \& Ostriker 2003). 
The UV output from such quasars, if not absorbed heavily by the surrounding torii, 
can contribute both to reionization and to the IR flux as viewed today. The unresolved 
soft X-ray background at energies between 0.5 and 2 keV may rule out a large density 
of miniquasars at high redshifts as required to reionize the universe from the quasar 
emission alone (Dijkstra, Haiman \& Loeb 2004) unless reionization by miniquasars, within 
uncertainties of models used in Dijkstra et al., is completed by a redshift of 20.
Here, we suggest that if all massive stars are eventually converted to black holes 
and accrete efficiently at the Eddington-rate (as would be the case
if miniquasars are responsible for eventual reionization of the Universe), then the IR 
background is overproduced. A more realistic situation could be that the mass 
function is broad, and only a small fraction of the total initial stellar mass is 
converted to black holes; current understanding of the IRB may limit 
the mass fraction of stars above 260 M$_{\sun}$ to a level less than 10\%. 
This constraint can be relaxed if miniquasars radiate less efficiently below the 
Eddington-limit; such a scenario may be viable if stellar sources dominate the UV photon 
production and are, thus, responsible for reionization, though some fraction of
stellar mass is converted to miniquasars.

The initial mass function of the first stars could potentially be 
dominated by stars with masses in the range between 140 and 260 $\msun$. 
In this case, the final product will be a complete disruption of the star via 
the pair-instability process (Heger \& Woosley 2002). In the case of such massive 
stars, in general, one expects a typical supernovae to be much brighter than 
the star, suggesting that a modest contribution to the IRB could come from 
supernovae, when compared to the fractional contribution from stellar 
emission alone. However, though supernovae are in fact brighter than an 
average star, the duration over which most optical flux is emitted is 
relatively smaller. We show that this results in a small but non-negligible 
contribution to the IRB from supernovae associated with massive Pop~III stars. 
While the background contribution is smaller, these supernovae are likely 
to be detectable directly in deep IR images above 1 $\mu$m 
down to AB-magnitude limits at the level of 26 and fainter.

In the next section, we discuss the contribution to IRB from the first generation 
of stars, supernovae, and miniquasars under the assumption that the early 
stellar population is dominated by very massive stars. In the case of supernovae, 
we will suggest that, while the relative contribution to the IRB may be lower, 
individual detections may be possible with deep-IR imaging. We also suggest that 
the IR background itself may be used to constrain the presence of early mini-quasars.
Throughout the paper, we make use of a $\Lambda$CDM cosmological model consistent 
with current data (e.g., Tegmark et al. 2004)
with $\Omega_m=1-\Omega_\Lambda=0.3$, $h=0.7$, $\Omega_b=0.04$,
the spectral index of the primordial power $n=1$, and 
a normalization for the matter power spectrum with $\sigma_8=0.9$.

\begin{figure}
\centerline{\psfig{file=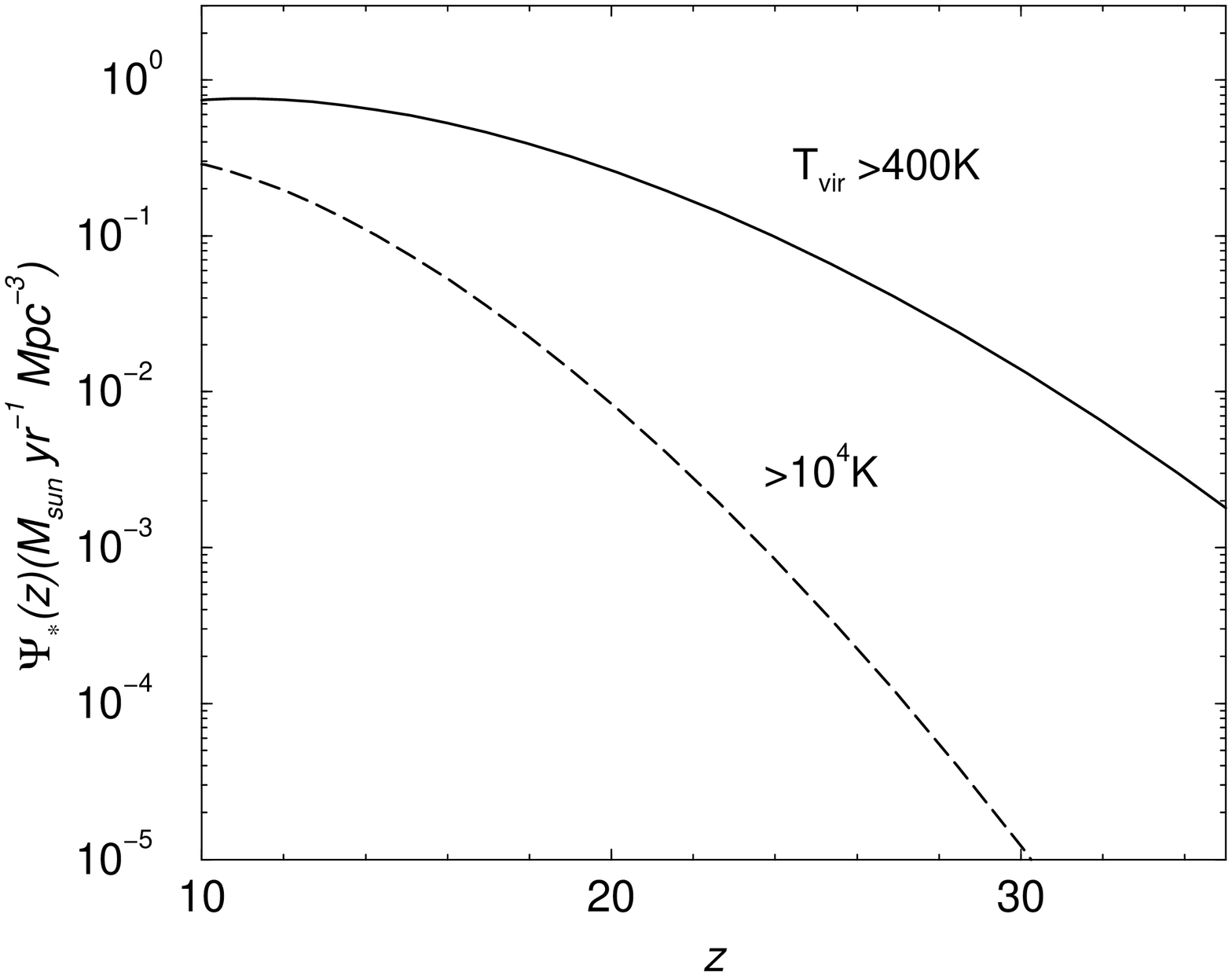,width=3in,angle=-90}}
\centerline{\psfig{file=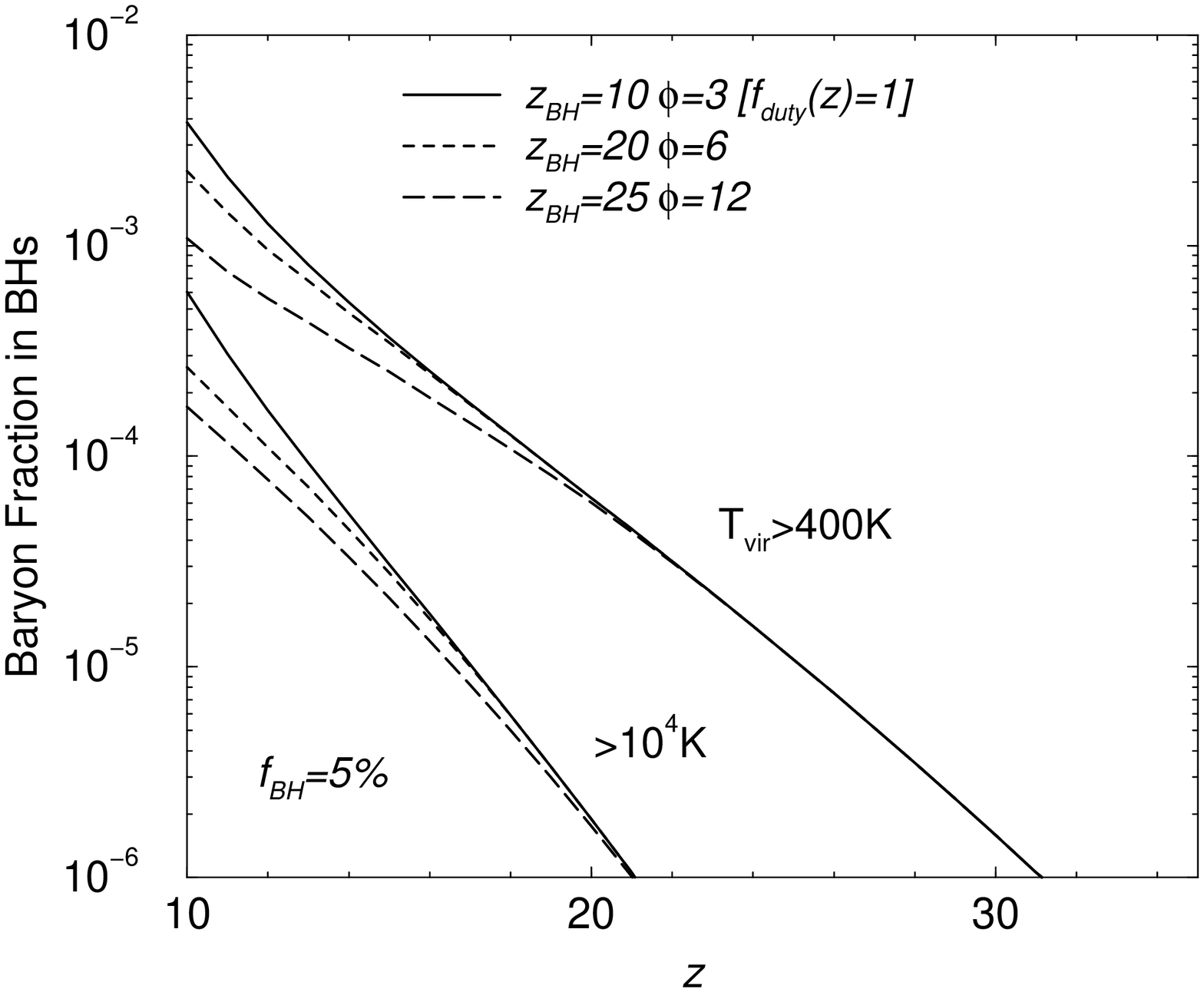,width=3in,angle=-90}}
\caption{{\it Top panel:}
The global star formation rate density at redshifts greater than 10 
calculated based on the Press-Schechter (PS) mass function with a 
minimum virial temperature of 400 K (top line) and 10$^4$ K (bottom line). 
{\it Bottom panel:} The baryon fraction in black holes as a function 
of redshift, under the assumption that the star-formation history
related to the seed population follows the two curves shown in the 
top panel. We have assumed a 5\% of the mass in stars are converted 
to seed black holes. The three curves in each of the star-formation models 
follow a model description related to the duty cycle of accretion growth 
following Ricotti \& Ostriker (2003) with parameters shown on the figure 
(see Section~2.1 for a discussion).}
\label{fig:sfr}
\end{figure}

\begin{figure}
\centerline{\psfig{file=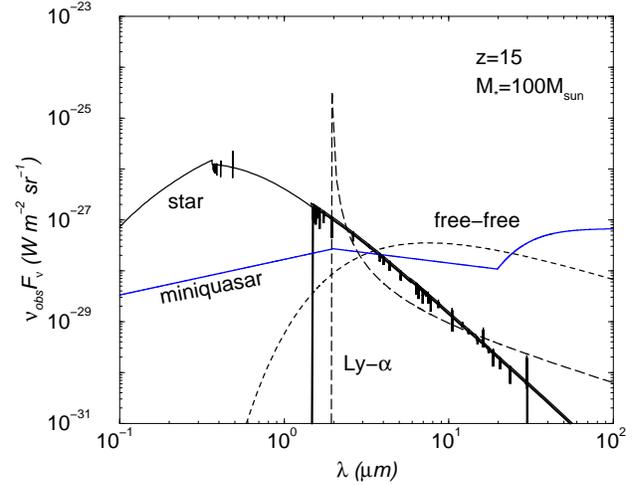,width=3.2in,angle=-90}}
\caption{The flux spectrum, $\nu F_\nu$, of a 100 $M_{\odot}$ star at a redshift 
of 15 as observed today as a function of the observed wavelength.
In addition to the stellar spectrum (solid line), we also show the 
nebular Lyman-$\alpha$ emission (long-dashed line) and free-free 
(dashed line) of the ionized H{\sc ii} region surrounding the star, when the 
UV end of the stellar-spectrum (thin solid line) is used to ionized the 
surrounding medium. The spectrum follows from model calculations by Santos et al. 
(2002). We assume all ionizing photons are absorbed by the nebula. For comparison, 
we also show the spectrum of a miniquasar of the same mass ($M_{\rm BH,seed}=100M_{\odot}$)
at a redshift of 15, as viewed today (solid line labeled 'miniquasar'). 
The spectrum is computed using the template derived in Sazonov et al. (2004). 
The bump around $\sim$ 100$\mu$m shows a possible contribution
from thermal reradiation of UV photons by dust.
While the overall flux from a miniquasar is similar to that of a star, 
we argue that the contribution to IRB from miniquasars could dominate stars 
because of their larger number density; quasars live longer than massive stars.}
\label{fig:star}
\end{figure}

\section{First-Sources in the IR Background}

Our calculations of the first-source contribution to the IRB follow 
previous calculations in the literature related to the stellar contribution 
to the IRB (Santos et al. 2002; Salvaterra \& Ferrara 2003; 
Cooray et al. 2004; Kashlinsky et al. 2004). 
These studies involve two basic ingredients: 
(1) the rate at which the volume density of first-sources evolves and 
(2) the flux spectrum.  We extend these calculations to supernovae
and miniquasars, which have been so far ignored in previous estimates,
under the assumption that 
(1) the supernova rate in the early universe traces that of the star 
formation and 
(2) the formation rate of miniquasars is simply related to the formation 
rate of seed black holes, taken to be proportional to the star-formation,
and to a model description for accretion. We use a simple model to describe 
the spectrum and the light curve of supernovae following results from 
Heger et al. (2002). 

For miniquasars, we make use of the average spectrum of quasars derived 
by Sazonov et al. (2004) using fits to observational data over the range
of wavelengths corresponding to X-rays and optical and, thereafter, using 
models to describe the emission, such as in the IR band. While the spectral 
shape is well defined in the average sense for observed luminous quasars at 
redshifts less than $\sim$ 6,  it is unclear to which extent this spectrum can be 
applied to miniquasars at redshifts greater than 10 and with masses in the 
range of 10$^2$ M$_{\sun}$ to 10$^4$ M$_{\sun}$.
At energies below 13.6 eV, the Sazonov et al. (2004) spectrum, $\nu F_\nu$, 
scales as $\nu^{-0.7}$ though expectations for miniquasars are that the 
spectrum is harder such that $\nu F_\nu$ is a constant (Madau et al. 2004; 
Dijkstra, Haiman \& Loeb 2004). Regardless of the exact shape of the spectrum, 
there is also an uncertainty with respect to the overall normalization, or the 
total luminosity of a given miniquasar. Here, we make the assumption 
that miniquasars radiate efficiently at the maximum allowed by the Eddington-limit. 
In making this choice, we are guided by observations of quasars out to redshifts of 6 which 
indicate maximal emission at the Eddington-limit. These quasars are more massive 
and more luminous than the miniquasar counterparts at high redshifts. 
It is likely that the miniquasar emission is submaximal. The limits we derive 
here on the fraction on miniquasars present during the reionization era should, 
thus, only be taken as a conservative limit on the low end given the current 
state of knowledge on the IR background.

\subsection{Formation rates and background flux}

In order to calculate the formation rate of first stars and the subsequent 
supernovae or seed black holes, we use an analytical description for 
halo formation following the Press-Schechter formalism 
(Press \& Schechter 1974). 
We write the star-formation rate as
\begin{equation}
\psi_*(z) = \eta \frac{\Omega_b}{\Omega_m} \;
\frac{\ed}{\ed t} \int_{M_{\rm min}}^\infty \ed M\; M \frac{\ed n}{\ed M} \, ,
\end{equation}
where $M_{\rm min}$ is the minimum mass of halos in which gas can cool, 
and we take the star-formation efficiency $\eta=0.4$. 
We consider two cases for $M_{\rm min}$ with a minimum virial
temperature of 400 K, involving molecular hydrogen (e.g., Tegmark et al. 1997), 
and 10$^4$K involving atomic hydrogen line cooling (Barkana \& Loeb 2001). 
The respective star-formation rates are shown in Fig.~1 (top panel). 
We mention that, although more detailed modeling of early star formation 
is available (e.g. Hernquist \& Springel 2003; Yoshida et al. 2003), 
the overall feedback effects from mini-quasars on star-formation 
are still uncertain (see, e.g. Machacek, Bryan, Abel 2003). 
We thus prefer using the model described above for the sake of simplicity. 
Note that the scenario with a minimum temperature $T_{\rm vir}=400$K likely provides 
a possible maximal star-formation rate.

Given the rate of formation of stars, we can write the emissivity per comoving 
unit volume at a certain wavelength $\lambda$ as a function of redshift $z$ as
\begin{equation}
j^c_\nu(z) =  \frac{l_\nu}{4\pi}\;  \langle t_{\rm age} \rangle \; \psi_*(z)  \, ,
\end{equation}
where $l_\nu$ is the luminosity per source mass as a function of frequency and 
$t_{\rm age}$ is the lifetime over which this flux is emitted; for a single 
source at a redshift of $z$ with a source mass $M_s$,  such as a first-star,
the total Luminosity, as a function of frequency, is $L_\nu = l_\nu M_s$ and 
the observed flux today is $F_\nu = L_\nu(1+z)/[4\pi d_L^2]$, where $d_L$ is 
the cosmological luminosity distance out to a redshift of $z$.

The cumulative background is obtained by integrating the emissivity over 
redshift, yielding the specific intensity $I_\nu$
\begin{equation}
v_{\rm obs}I_\nu = c \int_0^\infty \ed z \frac{\ed r}{\ed z} \nu(z) 
\frac{j_\nu^c(z)}{1+z} \, ,
\label{eqn:back}
\end{equation}
where $v_{\rm obs}$ is the observed frequency, $\nu(z)=(1+z)v_{\rm obs}$ is 
the redshift scaling of the frequency, and $r$ is the proper distance such that
$dr/dz = 1/[(1+z)H(z)]$ when the expansion rate for adiabatic cold dark matter 
cosmological models with a cosmological constant and a flat space-time geometry 
is $H^2(z) = H^2_0 [\Omega_m(1+z)^3 +\Omega_\Lambda]$.  
In terms of the individual source fluxes, and their comoving number density 
$n_s(z)$, the specific intensity of the background can also be written as 
$\nu_{\rm obs}I_\nu = \int \ed z\; (\ed V/\ed\Omega dz)\; \nu(z) F_\nu(z) n_s(z)$,
where $n_s(z) =   \langle t_{\rm age} \rangle \; \psi_*(z)/M_s$ 
and $(\ed V/\ed\Omega\ed z)$ 
is the cosmological volume element given by $d_L^2/(1+z)^2\; \ed r/\ed z$, in 
terms of the luminosity distance. This expression and equation (\ref{eqn:back}) 
are equivalent.

We use the same formalism to compute the background radiation 
intensities from stars, supernovae and miniquasars. We assume that the 
supernovae rate follows that of the star-formation. 
For miniquasars, we follow Ricotti \& Ostriker (2003) and write the
growth rate of black holes as
\begin{equation}
\dot \omega_{\rm BH}(z) = \dot \omega_{\rm ac}(z) + \dot \omega_{\rm seed}(z) \, ,
\end{equation}
where the two terms represent the growth of black holes via accretion,
$\dot \omega_{\rm ac}$, and the formation rate of seed black holes, 
$\dot \omega_{\rm seed}$. The over-dot represents the derivative with 
respect to proper time. We ignore black hole removal process, such as 
through ejection from proto-galaxies, and take 
$\dot \omega_{\rm seed}(z)=f_{\rm BH}\psi_*(z)$ 
where $\psi_*(z)$, the star-formation rate, is given by equation (1),
and $f_{\rm BH}$ represents the overall fraction of stars converted to 
seed black holes.  We model growth related to accretion as 
$\dot \omega_{\rm ac}(z)=f_{\rm duty}(z)\omega_{\rm BH}(z)/t_{\rm Edd}$, 
where Eddington time is set to be a constant at 10$^8$ years and use 
the same parametrization in Ricotti \& Ostriker (2003) for 
$f_{\rm duty}$ such that $f_{\rm duty}(z)=[(1+z)/z_{\rm BH}]^\phi$ with 
the condition that
$f_{\rm duty}(z) \leq 1$ based on the definition that 
$f_{\rm duty} = t_{\rm on}/(t_{\rm on}+t_{\rm off})$ where
$t_{\rm on}$ and $t_{\rm off}$ are time intervals over which the blackhole 
accretes and does not accrete, respectively.
At the low end, following Ricotti \& Ostriker (2003), we set 
$f_{\rm duty}(z) > 10^{-3}$ to be consistent with both the AGN fraction at 
$z \sim 3$ and today (Steidel et al. 2002). In Fig.~1 (bottom panel), 
we show $\omega_{\rm BH}(z)$ normalized to the baryon density and 
assuming $f_{\rm BH}=0.05$ for a variety of models related to 
$f_{\rm duty}$ with parameters as labeled on the figure.

\subsection{Source Spectra}

The spectrum of stellar emission is computed as in Santos et al. (2002), 
including the nebular and free-free emission. For the miniquasar spectrum, 
we make use of the model of Sazonov et al. (2004). We assume that the 
Sazanov et al.'s template for the average quasars also applies to 
miniquasars. In Fig.~2, we compare the spectrum of a typical star 
with $100M_{\odot}$ and that of a miniquasar at a redshift of 15, 
as viewed today. Here we plot $\nu_{\rm obs}F_\nu$ (in units of W m$^{-2}$
sr$^{-1}$). The miniquasars are assumed to radiate at the Eddington-limit 
such that the total integrated luminosity over the range from X-rays to 
far-infrared is $\approx 1.3\times10^{38}$ ergs s$^{-1}$ (M$_{\rm BH}/\msun$), 
though, as discussed in Section 2, it is likely that the miniquasar emission 
is submaximal. Also, at observed wavelengths below $\sim$ 2 $\mu$m, corresponding 
to redshifted Lyman-$\alpha$ emission, the miniquasars flux spectrum $\nu F_\nu$ 
scales as $\nu^{-0.7}$. For miniquasars, it could be that the spectrum is harder 
(in terms of energy), such that the expectation is that $\nu F_\nu$ is constant 
at energies above 13.6 eV (Madau et al. 2004; Dijkstra, Haiman \& Loeb 2004). 
This will result in a flatter spectrum, as observed today, from optical/UV to IR 
wavelengths.
As shown in Fig.~2, we assume that stellar emission is responsible for 
reionization of the universe; the Lyman-$\alpha$ emission is related to ionizing 
photons that are first absorbed by the neutral medium and are re-emitted during 
recombinations. In the case of miniquasars, we do not consider UV absorption
as we have implicitly assumed that the surrounding medium is already ionized by 
stars preceding the formation of a miniquasar in or near that location. 
We will comment on this later in the discussion.
As shown in Fig.~2, the level of the incident flux from a typical star and a 
miniquasar is similar. 

The spectrum and light curve for Population~III supernovae were calculated by 
Heger et al. (2002). In these calculations, the relevant IR 
emission, as observed today, characterizing the peak of the light curve, is delayed 
a month or so from the initial explosion and the resulting shock break out. 
The latter includes emission at short wavelengths, below the Lyman limit,  
in the supernova frame, which is likely be absorbed by the intergalactic medium at 
epochs prior to complete reionization. The peak of the light curve extends over 
a month at most. In order to make a reasonable estimate of the supernovae
contribution to the IRB, we use the peak emission and its duration only and ignore 
detailed aspects of the light curve when fluxes drop below more than a factor of 10 
from the peak emission. Including the full light curve only leads to minor 
corrections at a a few percent level. While the peak emission is flat such that 
$\lambda F_\lambda$ is a constant in the estimate by Heger et al. (2002), to study 
any departures, we also allow the spectrum to vary as 
$\lambda F_\lambda \propto \lambda^\alpha$ with $\alpha$ between -0.5 and 0.5, 
with $\alpha=0$ as the fiducial case. 
When varying the spectral shape, we renormalize such that the total flux, or the 
luminosity, remains constant in the rest wavelengths between 10$^2$ $\AA$ and 
10$^4$ $\AA$ corresponding to the UV and optical regimes.
Here, again, we assume that the surrounding medium is reionized by stellar emission 
before the supernova explosion and that the rest UV emission is not absorbed. 
The reionization by supernovae may not be important as the time scale over
which the rest UV emission is expected is significantly smaller than that of a star 
(a month vs. a few million years, respectively).
Note that the supernova spectrum of a Pop III star is highly uncertain 
in terms of spectral shape at the UV end and more numerical models are needed 
to further address the extent to which supernovae may be an important source of 
ionizing photons. A more likely scenario involving Pop III supernovae is the one 
described in Oh, Cooray \& Kamionkowski (2003), where the blast wave propagates to 
the dense ionized IGM and heats the electrons to a substantial temperature which 
subsequently cools via inverse Compton-scattering off of the cosmic microwave 
background, similar to the same effect related to hot electrons
in galaxy clusters (the Sunyaev-Zel'dovich effect; Sunyaev \& Zel'dovich 1980). 

\begin{figure*}
\centerline{\psfig{file=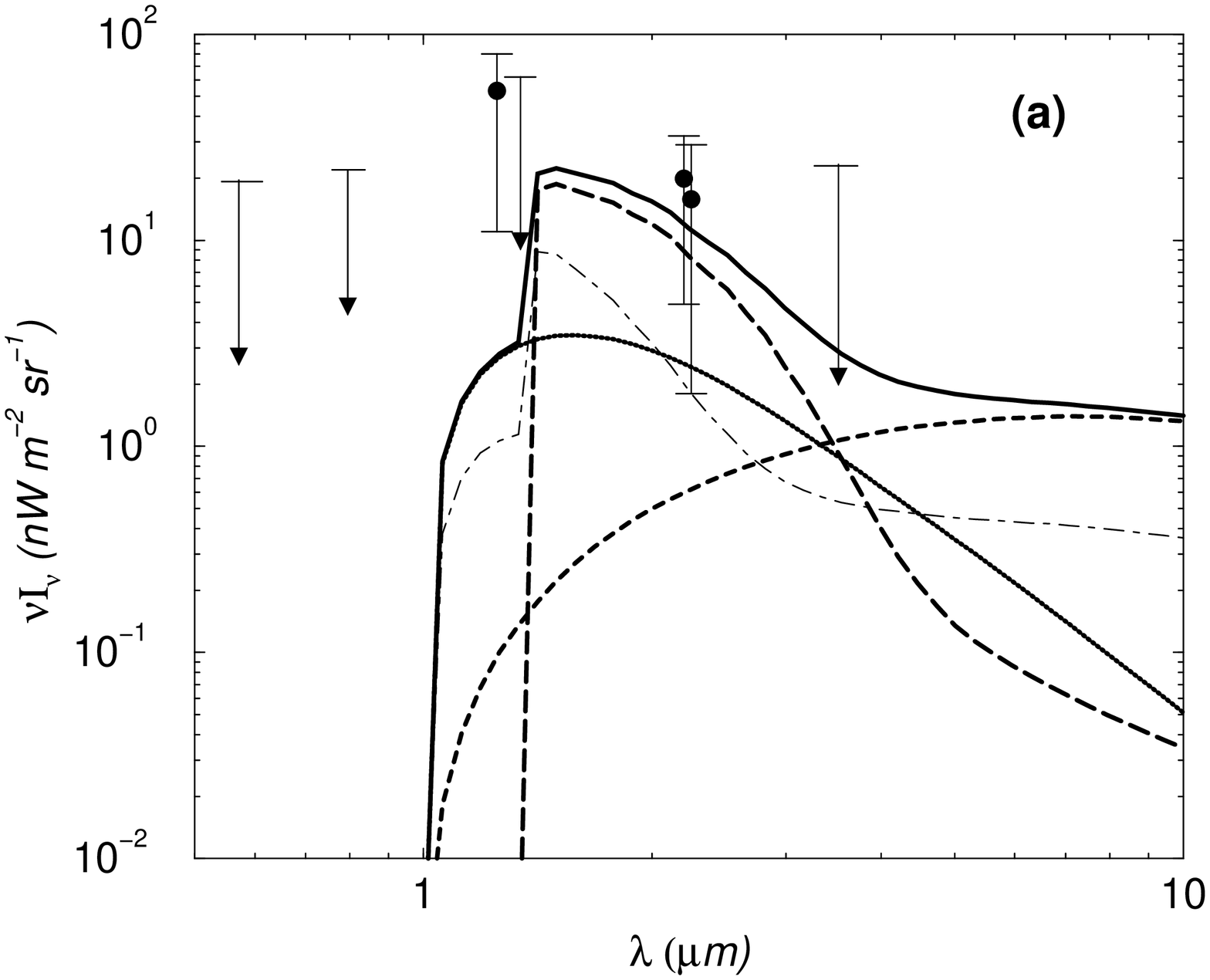,width=2.3in,angle=-90}
\psfig{file=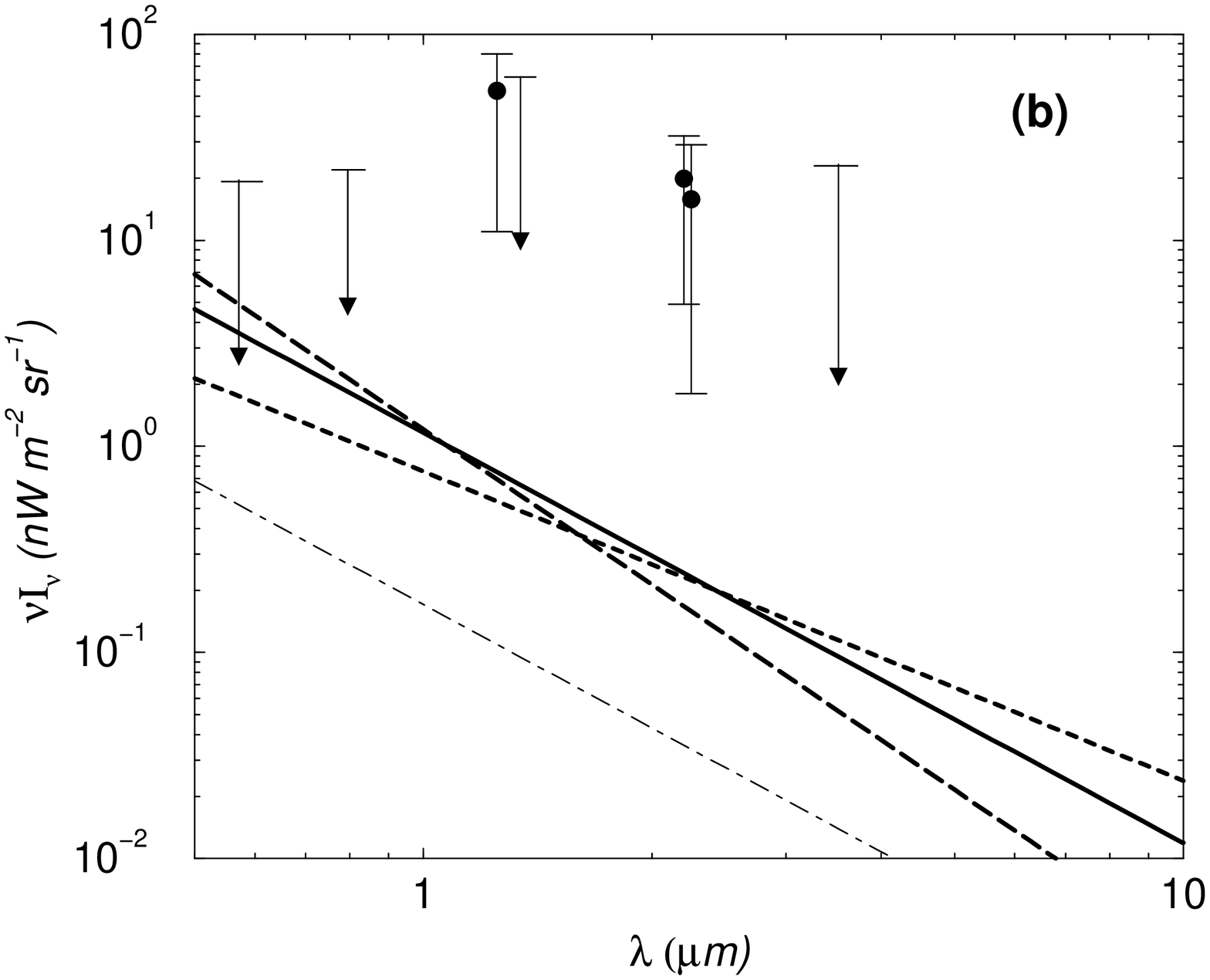,width=2.3in,angle=-90}
\psfig{file=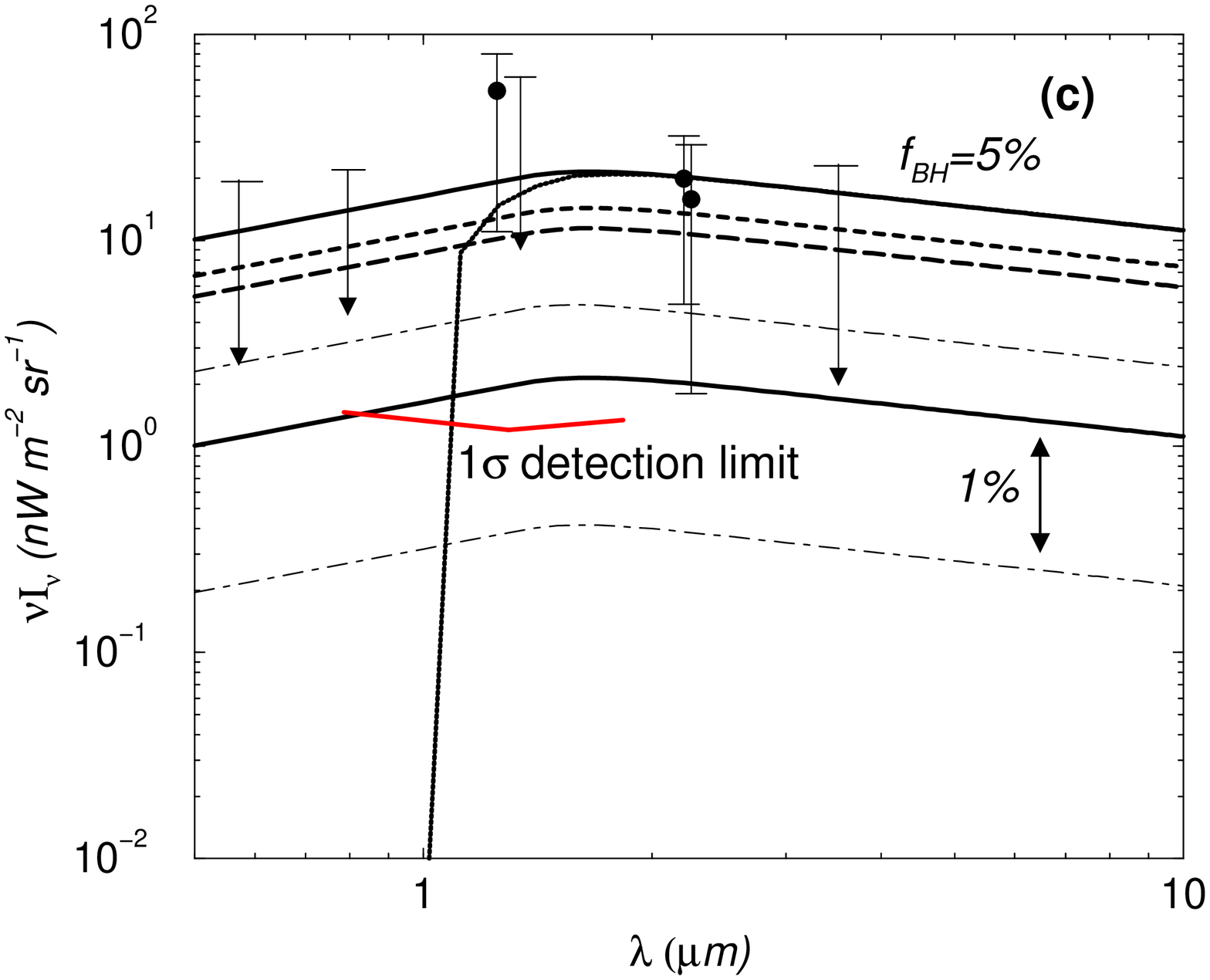,width=2.3in,angle=-90}}
\caption{The cosmic infrared background due to Population III stars (a),
first supernovae (b), and miniquasars (c). 
The data points and upper limits are related to unexplained IRB intensities 
from measurements and calculations in the literature. For reference, 
we show the same observational data as the ones plotted in Fig.~6 of Santos et al. 
(2002; the measurements at 1.25 $\mu$m and 2.2 $\mu$m from Cambr\'esy et al. 2001; 
upper limit at 1.25 $\mu$m, a second measurement at
2.2 $\mu$m, and a upper limit at 3.5 $\mu$m from Wright \& Johnson (2001); 
upper limits at 0.6 $\mu$m and 0.8 $\mu$m from Bernstein, Freedman \& Madore 2002). 
Note that these background measurements are corrected for the known contribution 
from Galactic stars and galaxies down to magnitude limits of order 25 
(see, Santos et al. 2002 for further details).
In the case of stellar contribution to the IRB shown in Panel (a), 
we show the contribution from stellar emission alone (dotted line),
associated Lyman-$\alpha$ emission due to recombinations in the ionized 
patches surrounding individual stars (long-dashed line),
the free-free emission due to electron/ion scattering within individual 
ionized patches (dashed line), and the sum of these
three contributions (thick solid line). These curves make use of the
star formation rate (SFR) density based on the PS mass function with a minimum 
temperature for collapse of 400 K; for illustration, the bottom dot-dashed curve 
is the total contribution to the IR background in the case where minimum temperature 
is set at 10$^4$ K. In the case of supernovae, in Panel (b), 
the three curves assume flux spectra for supernovae emission, 
$\lambda F_\lambda \propto \lambda^\alpha$ 
with $\alpha=0$ (solid line),-0.5 (dashed line),0.5 (long-dashed line); The curves 
are normalized such that the total flux or total luminosity is same in the UV to 
optical bands (see Section~2.2 for details). 
The top curves assume PS mass function to calculate the SFR with T$_{\rm vir}$=400K, 
but the case with T$_{\rm vir}$=10$^4$K, for $\alpha=0$, is shown as a dot-dashed line.
In the case of miniquasars in Panel (c), we consider three models for the accretion duty 
cycle as plotted in Fig.1 (bottom panel) with parameters given in that figure and 
plotted as top three curves with thick lines when T$_{\rm vir}$=400 K. The middle 
dot-dashed curve is for the case corresponding to the solid line but with T$_{\rm vir}$=10$^4$K.
These curves assume a seed black hole mass fraction, $f_{\rm BH}$, of 5\%. For reference, 
we also show the range implied by the solid line and the dot-dashed line, with thin 
bottom lines, when this fraction is reduced to 1\%. Note that we have assumed the
case that no UV photons are absorbed by the IGM surrounding these miniquasars. The 
solid curve labeled '1 $\sigma$ detection limit' is the extent to which the optical to near 
IR background can be established, over the wavelength regime indicated by the curve, using 
a low resolution spectrometer on a proposed rocket experiment.}
\label{fig:irb}
\end{figure*}

\subsection{Infrared background}

We summarize our results in Fig.~3, where we show the contributions to 
total IRB light from stars, supernovae and miniquasars. 
In the case of stars, we divide the contribution to the three main 
components; direct emission from stars, Lyman-$\alpha$ recombination
radiation from the ionized patch, and the free-free component related 
to electron-ion scatterings within each ionized patch
surrounding these stars. Our calculations related to the IRB light due 
to stars are consistent with those in Santos et al. (2002) and 
Salvaterra \& Ferrara (2003), except that here we have assumed a slightly
different redshift limit to the first generation of stars at a value of 10 
instead of values such as 7 and 8 used in those calculations that are
aimed at explaining the total missing content. 

Our results suggest that, while the emission associated with stars can explain 
most of the IRB intensities, the fractional contribution from supernovae 
is an order of magnitude smaller than the stellar emission contribution. 
The relative difference between stars and supernovae is easily understood 
based on the typical flux from a star vs. a supernovae and the ages 
over which these fluxes are emitted.
For example, typical $\sim$ 200 M$_{\odot}$ star at a redshift of 20 has 
a flux, ignoring Lyman-$\alpha$ emission, of order 
$\sim$ 10$^{-35}$ ergs cm$^{-2}$ s$^{-1}$ Hz$^{-1}$ ($\sim 10^{-3}$ nJy or 
$m_{\rm AB} \sim 39$), 
while the peak flux of the supernovae at the same redshift is at the level of 
$\sim$ few times 10$^{-30}$ ergs cm$^{-2}$
s$^{-1}$ Hz$^{-1}$ ($\sim$ 1 $\mu$Jy or $\sim$ $m_{\rm AB}\sim 26$). 
Even if all the stars at redshifts greater than 10 die as supernovae, 
though the instant flux is greater, the total output of $(F_\nu t_{\rm age})$ 
from supernovae is smaller than stars as the stellar emission lasts over 
a million years while the peak flux from supernovae lasts just over a month or so. 
Thus, the apparent high ratio of the 
{\it peak} luminosity is largely compensated by the age difference so that 
the typical ratio of a supernova to stellar flux remains at most at the level 
of $\sim$ 0.1. 

Note that the stellar contribution to the IRB also includes a substantial 
fraction from the Lyman-$\alpha$ emission of absorbed ionizing UV photons during 
subsequent recombinations. We have not included such a contribution from individual 
Pop III supernovae because of uncertainty in the supernovae spectrum at the UV end. 
Note also that regions surrounding supernovae is likely to have been ionized by
the progenitor stars. 
Even if the IGM between proto-halos are partly ionized, the resulting Lyman-$\alpha$ 
emission is expected to be suppressed relative to the case of stellar sources. 
This can be understood from the fact that the Lyman-$\alpha$ spectrum, as a function 
of frequency, can be written as 
$l_{\rm Ly\alpha}(\nu) =q h \; \nu_{\rm Ly\alpha} \phi(\nu)$,
where $\phi(\nu)$ is the fit to the Lyman-$\alpha$ profile of Loeb \& Rybicki (1999) 
by Santos et al. (2002) and the production rate of Lyman-$\alpha$ photons from 
recombinations is $q \sim 0.6 \langle t_{\rm age} \rangle \alpha_B(T) n_e^2 \dot{R}$, 
where $n_e$ is the electron density of the surrounding medium,
$\alpha_B(T)$ is the case-B recombination coefficient for HI as a function of the 
electron temperature $T$, and $\dot{R}$ is the ionizing photon production rate 
(see Santos et al. 2002 for details).  To estimate $\dot{R}$, we assume that the 
supernovae spectrum is flat (in rest $\nu F_\nu$ below the Lyman-limit) and given the
uncertainty in spectral shape, we take the production rate to be that associated with
ionizing photons at the wavelength corresponding to the Lyman-limit. 
The model related to stars is shown in Fig.~2, where the unabsorbed spectrum peaks 
at a wavelength of $\sim$ 300 $\AA$ (rest frame).
The difference in flux, as suggested above, translates to a stellar-to-supernovae 
ionizing photon production rate difference of $\sim$ 10$^{-5}$ to 10$^{-6}$. This 
should then be compared to the ratio of ages, which is again, $\sim 10^7$ between 
stars and supernovae. Furthermore, in the case of supernovae, the Lyman-$\alpha$ 
radiation is more likely to be associated with tenuous IGM with density 
$\sim 10^{-7} (1+z)^3$ cm$^{-3}$ compared to dense nebulae surrounding stars 
(with density $10^4$ cm$^{-3}$). This lowers the Lyman-$\alpha$ flux by
another factor of 10$^6$ relative to that of stars, resulting in an overall 
reduction of 10$^{-5}$ to 10$^{-7}$ in the Lyman-$\alpha$ contribution from stars 
to supernovae.  Thus, it is unlikely that the difference between
predicted supernovae background and the missing IRB flux can be reconciled 
with the expected increase associated with Lyman-$\alpha$ emission if the UV emission 
from supernovae are absorbed during the reionization process.

Intriguingly, redshifted light from early miniquasars can explain 
the IR excess at wavelengths around 1 $\mu$m, as shown in Fig~3(c).
In fact, if {\it all} stars collapse to become seed black holes, 
we find that the flux from miniquasars that grow down to $z\sim 10$ will 
{\it exceed} the measured IR fluxes by an order of magnitude or more. 
A fractional contribution of $\sim 5\%$ to 10\% of initial stellar mass 
to seed black holes appears consistent with observations. This is based on our
assumption that the miniquasars are radiating efficiently at the maximum given by 
the Eddington-limit. As discussed at the beginning of Section~2, the extent to which 
such an assumption, based on observations of luminous quasars at
redshifts less than 6, applies to high-redshift miniquasars is unclear. 
If the radiation from miniquasars, with masses between 10$^2$ M$_{\sun}$ and 
10$^4$ M$_{\sun}$, is submaximal, the derived limit on the fractional contribution 
from $\sim 5\%$  to 10\% can be increased. Thus our result should be considered 
as a {\it weak} constraint.

Since the typical stellar age is $\sim 2 \times 10^6$ years 
(Bromm, Kudritzki \& Loeb 2001), at a redshift of around 10, using the highest 
star-formation rate in Fig.~1 (top panel), one finds a typical density of order 
a few times $10^6$  $\msun$ Mpc$^{-3}$ in stars. From the bottom panel of Fig.~1, 
one finds that the density of miniquasars is typically $\sim 10^8$ $\msun$ Mpc$^{-3}$.
This apparent increase in miniquasar density, however, is compensated by the 
fact that the miniquasars flux, at relevant wavelengths, is roughly an order 
of magnitude smaller than that of a typical very massive star (see Fig.~2). 
The exact numerical calculation reveals that the miniquasars do dominate the 
background if the fraction of stars converted to seed black holes is large. 
In other words, the IRB offers a useful way to
constrain the presence and abundance of miniquasars at $z>10$;
under our assumption, an initial mass fraction of black holes, relative to stars, 
of order 10\% and below seems consistent with current observations.

This conclusion may indeed be strengthened if ionizing photons from miniquasars are 
absorbed by the neutral IGM; if absorbed, we expect some fraction of the 
energy to be reradiated in the form of the Lyman-$\alpha$ line
and to increase the IRB at the corresponding frequencies such that the upper 
bound on the black hole mass fration is further reduced. 
On the other hand, as discussed in the case for supernovae above, we do not 
expect the Lyman-$\alpha$ radiation to be significant
as long as the dense nebula surrounding quasar is fully ionized already by the 
stellar emission prior to the formation of the miniquasar. The miniquasar emission 
can ionize the IGM beyond the nebula, but due to the lower density, recombinations 
are unlikely to be significant such that the Lyman-$\alpha$ emission is suppressed 
relative to the case where stars form and ionize dense nebulae surrounding such 
stars initially.

Our suggestion that the seed mass fraction must be below 10\%, assuming 
efficient radiation at the Eddington-limit, also suggests that the universe cannot 
be reionized by early quasars alone. For reference, in Fig.~3, we show the model 
in which the UV photons from miniquasars are absorbed by neutral IGM during the 
reionization process (thick dot-dashed line in the figure).
Integrating the UV light below the hydrogen Lyman-limit, 
we find a ratio of number density of photons relative to mean density of 
baryons at the level of 0.3 at $z \sim 10$; to fully reionize, one would expect 
this ratio to be at least unity or as high as 25 as the recombinations are
important at high redshifts. In Oh, Cooray \& Kamionkowski (2003), it was estimated 
that $25 \pm 12 (f_{\rm esc}/0.3)^{-1} (\tau_e/0.12)
(C_{\rm II}/4)$ photons per baryon are needed to explain an optical depth to reionization 
of 0.12 when the clumping factor of ionized gas $C_{\rm II}$ is 4 and the escape fraction 
of ionized photons is 0.3, consistent with other estimates 
(e.g., Dijkstra, Haiman \& Loeb 2004). Note also that ionization by a combination of 
hard X-rays and secondary electrons does not lead to complete ionization, as argued 
by Madau et al. (2004).

Our constraint on the total mass fraction of miniquasar-seed blackholes is independent 
of similar suggestions in the literature, for example, based on the unresolved X-ray 
background (Dijkstra, Haiman \& Loeb 2004). These authors argue that
the X-ray background would be violated by a population of miniquasars if 
they are responsible for complete reionization. If we consider a scenario where 
miniquasars exist with stars, with a seed miniquasar mass fraction of 10\%,
we find that the X-ray background constraint at 1 keV is not violated by this miniquasar 
population even in the extreme case where the spectrum is normalized to the
Eddington-limit. Interestingly, the level of the IRB background from the same population 
is consistent with current observations. Given the uncertainties in the available 
observations and in our analytical model, we are not yet able to exactly establish the 
allowed seed mass fraction.

Although current observations of the IRB are still uncertain, one can improve the limit 
related to the mass fraction of seed black holes further by future background 
measurements. In Fig.~3(c), we show the 1$\sigma$ noise, or detection, level related 
to a planned low-resolution spectroscopic background measurement on a rocket
experiment (T. Matsumoto, private communication) for IRB fluctuation 
studies (e.g., Magliocchetti et al. 2003; Cooray et al. 2004). The detection limit comes 
from the addition of 400 spectra over a 30 second integration time at an altitude 
above 300 km (which removes the air-glow) and after a foreground
source subtraction down to 14, 15.9 and 16.3 magnitudes in I-, J- and K-bands. 
The background can be measured to a level of 1 nW m$^{-2}$ sr$^{-1}$, whereas the
limiting factor in these measurements will be related to the removal of the zodiacal light 
(e.g., Cambr\'esy et al. 2001). In this case of the rocket experiment, this removal is 
achieved with data taken as a function of the latitude such that, when combined with 
multifrequency aspect, removal down to the instrumental noise level is expected. 
The proposed measurement would help establish the excess above Galactic stars and 
foreground sources better than what is currently known and, in return,
limit the extent to which high-z sources, either in the 
form of stars and/or miniquasars, may be present.

\subsection{Far-Infrared background}
One may consider the possibility whether the constraint from the IRB can be relaxed if the rest UV 
and optical emission from miniquasars is heavily absorbed by dust. First, we note that the
extinction by inter-galactic dust is unimportant at $z>10$ even if
nearly all the first stars die as pair-instability supernovae 
(see Yoshida et al. 2004). If the inter-stellar absorption due to dust within (proto-)galaxies
that are hosting miniquasars is significant, the reradiation by thermal dust at far-IR 
wavelengths at $\sim 100 \mu$m and above can still be used to 
put limits on their abundance. Assuming the same spectrum as shown in Fig.~2 for quasars, 
which allows for significant dust absorption, we predict a 850 $\mu$m background of $\sim$ 
50 nW m$^{-2}$ sr$^{-1}$, for $f_{\rm BH}=5$\% while
the observed background is below a few nW m$^{-2}$ sr$^{-1}$ (Hauser \& Dwek 2001). 
This suggests that even for the 5\% seed mass case, which may explain the 
missing IRB, absorption by dust cannot be as significant as for low redshift
luminous quasars. Keeping the same seed mass fraction, one can refine the extent to which 
dust absorption is significant by studying the unresolved background at these far-IR 
wavelengths. While $\sim$ 40\% of the 850 $\mu$m and 350 $\mu$m backgrounds have now been 
resolved, the counts are such that a small extrapolation to a slightly lower flux density 
resolves the whole background (Borys et al. 2003).
Unlike the near-IR background where surface brightness converges down to AB magnitudes of 
$\sim 25$, as observed counts go deeper, such a convergence is not found at far-IR wavelengths.
Thus, it'll be useful to return to this aspect in the future when source counts are well 
defined. As an early conclusion, we are forced to consider the possibility that
the far-IR background can be fully explained with resolved source counts alone. 
This, when combined with IRB measurements, suggests that a substantial population of 
``dusty'' miniquasars is unlikely to be present at $z>10$.

\section{Summary}

The cosmic infrared background (IRB) at wavelengths between 1 $\mu$m and 
3 $\mu$m provides a useful probe of the global star-formation rate
prior to and during the reionization. We have studied the relative 
contributions from first stars, supernovae and early miniquasars to the 
infrared background (IRB). In addition to massive stars, the IRB flux may 
be dominated by a population of miniquasars, especially if the first 
generation of stars are very massive and the end product of the stars lead 
to a substantial population of black holes that will effectively radiate 
as miniquasars by matter accretion. 

We use the Press-Schechter formalism that describes dark halo formation, 
and combine it with a simple star-formation model to calculate the
cosmic star-formation history, the supernova rate, and the blackhole formation 
rate at high redshift. We follow previous calculations of the source spectrum
for stars, while for Population~III supernovae, we use a simple description of 
the peak emission and its duration. We describe the flux spectrum of 
miniquasars using an average spectrum as observed for luminous quasars at 
redshifts less than 5 based on observed data and models by Sazonov 
et al. (2004). We fix the freedom related to the overall normalization of this 
spectrum by assuming that the miniquasars radiate efficiently at the maximum 
allowed by the Eddington-limit. Under this assumption, we find that a mass 
fraction of seed black holes, relative to Pop~III stars, at the level of $\sim 10\%$ 
and below is consistent with observations. This upper limit on the seed mass fraction 
can naturally be extended to a higher value if miniquasars
radiate below the Eddington-limit, so the limit we derive should be considered as on 
the lower side rather than a strict upper limit.
Future spectroscopic background measurements exploiting rockets will 
improve the observational uncertainties and hence the constraints
on the high-redshift miniquasar population.

While the integrated light from Population III supernovae to the 
IRB is subdominant, one can potentially consider the possibility 
of directly detecting individual supernovae in high resolution deep IR 
imaging data. Given the star formation rate density and the expected flux, we 
estimate the surface density of supernovae to be order few tens 
per sqr. degree over a year, if the star-formation rate is close 
to 0.5 M$_{\odot}$ yr$^{-1}$ Mpc$^{-3}$ at $z\sim 10$, and assuming that all stars 
are massive ($ > 100$  M$_{\odot}$). For the high redshift supernovae, 
the typical AB magnitudes are at the level of 26 in 1.5 $\mu$m (Heger et al. 2002). 
These are within reach of deep IR imaging with existing large telescopes 
and from space and, in the long term, clearly within reach of missions 
such as the James Webb Space Telescope (JWST), which can detect 
point sources down to a magnitude limit of 31 around the same wavelengths 
in a 10$^4$ sec exposure in its 16 arcmin$^2$ field of view
\footnote{http://www.ngst.nasa.gov for sensitivities and other details}.

\vspace{0.5cm}

{\it Acknowledgments:} 
This work is supported by the Sherman Fairchild foundation and DOE DE-FG 03-92-ER40701 (AC). 
NY thanks support from Japan Society of Promotion of Science Special Research Fellowship (02674). 
We thank an anonymous referee for a careful reading of the manuscript
and for detailed comments that improved the discussion in this paper.

\end{document}